\documentclass[prc,superscriptaddress,amsfonts,aps,showpacs]{revtex4}
\usepackage{graphics,epsf,bm,color}
\def\he{$^4$He }
\def\hl{$^{\hspace*{0.1em}5}_\Lambda$He }
\def\hll{$^{\hspace*{0.6em}6}_{\Lambda\Lambda}$He }
\def\dll{$\Delta B_{\Lambda\Lambda}$}

\begin{document}
\title{Brueckner Rearrangement Effects in \hl and
$^{\hspace*{0.5em}6}_{\Lambda\Lambda}$He}
\author{M. Kohno}
\affiliation{Physics Division, Kyushu Dental College,
Kitakyushu 803-8580, Japan}
\author{Y. Fujiwara}
\affiliation{Department of Physics, Kyoto University,
Kyoto 606-8502, Japan}
\author{Y. Akaishi}
\affiliation{Institute of Particle and Nuclear Studies,
KEK, Tsukuba 305-0801, Japan}

\begin{abstract}
Rearrangement effects in light hypernuclei are investigated in the
framework of the Brueckner theory. We can estimate without detailed
numerical calculations that the energy of the $\alpha$-core is reduced
by more than 2.5 MeV when the $\Lambda$ adheres to \he to form \hl.
Similar assessment of rearrangement contributions is essential
to deduce the strength of $\Lambda\Lambda$ interaction from
experimentally observed $\Delta B_{\Lambda\Lambda}$. The recently
observed experimental value of $\sim$ 1 MeV for the
$\Delta B_{\Lambda\Lambda}$ of \hll suggests that the matrix element
of $\langle\Lambda \Lambda |v|\Lambda \Lambda\rangle$ in \hll is
around $-2$ MeV
\end{abstract}


\maketitle

\section{Introduction}
The hypernuclei serves as an invaluable source of information for
hyperon-nucleon and hyperon-hyperon interactions. New discovery of
the double hypernucleus \hll \cite{TAKA} provided more reliable
data which would revise previous understanding based on the old data
\cite{BLL1,BLL2,BLL3,BLL4} that the $\Lambda \Lambda$
interaction was fairly strong. A quantity \dll (\hll ), which is defined
as $B_{\Lambda\Lambda}(\mbox{\hll})-2 B_{\Lambda}(\mbox{\hl})$ or
equivalently $2E(\mbox{\hl}) -E(\mbox{\hll}) -E(\mbox{\he})$,
is meant to deduce the strength of the $\Lambda\Lambda$ interaction.
The new data indicated that \dll (\hll ) $\sim 1$ MeV. However, it
contains many body effects. Since exact few body calculations are not
practical yet apart from very light ordinary nuclei, the understanding
of these effects from the viewpoint of standard nuclear many body theory
is useful. It has also been pointed out that hypernuclear systems
work as an interesting testing ground of nuclear many body theories.
In this paper we estimate rearrangement effects \cite{BG} in a framework
of the Brueckner theory \cite{LOBT1,LOBT2} for \hl and \hll.
When matrix elements between nucleons and lambdas are needed,
we use the $SU(6)$ quark model potential \cite{FU01,FU02}, which provides
a successful unified description for octet baryon-baryon interactions.

Rearrangement effects in theoretical consideration of hypernuclei
have been discussed by many authors. Variational calculations \cite{BU,BMU}
with a Jastrow trial function addressed
the problem of the core polarization effect. The repulsive energy change
due to the nuclear core polarization was also discussed in mean field
calculations with a relativistic parameterization \cite{MLM} and
Skyrme type of density dependent effective forces \cite{LAN,CLS}.
Our discussion in this paper is focused on Brueckner rearrangement energies
which arise through the energy-dependence of the reaction matrix and the
Pauli principle. The correspondence of these effects to higher order
correlations in Fermi hypernetted chain approach is not straightforward.

We first present the treatment of the energy change of the \he core
in \hl from \he. This problem was discussed by Bando and Shimodaya \cite{BS80}
in relation with the overbinding problem of \hl. The estimation
of the rearrangement energy for the $\Lambda$ in nuclear matter was presented
by Dabrowski and K\"{o}hler \cite{DK} as early as in 1964.
In the present consideration
we show that numerical evaluation of relevant matrix elements can
be avoided and that the potential energy change is shown to be connected
to the $\Lambda$ separation energy of \hl and the wound integral of nucleon
pairs. Thus the estimation is more solid than that of Bando and Shimodaya.
Recently, Nemura, Akaishi and Suzuki \cite{NAS} showed by their variational
calculations that the addition of the $\Lambda$ decreases the \he core energy
in \hl by 4.7 MeV. The variational calculation in \cite{BU,BMU},
on the other hand, suggested that the rearrangement energy is small.
The discussion of the corresponding effects in the standard
nuclear many body theory is helpful.

Next we consider the rearrangement energy in  \dll (\hll ).
Again it is shown that the essential ingredient can be written by the $\Lambda$
separation energy of \hl, wound integrals of nucleon pairs and
nucleon-lambda pairs. In this case the Pauli rearrangement contribution
of the $\Lambda$ particle also appears.
The recognition of these possible rearrangement effects in hypernuclei
is important, because sophisticated calculations in which the structure
dependence of effective interactions is ignored have been often presented,
particularly, in cluster models

\section{Rearrangement effects in $^{\hspace*{0.1em}5}_\Lambda$He}

When one or two lambda particles are added
to $^4$He, the interaction between the $\Lambda$ and nucleons causes a
change of correlations among nucleons and thus the energy expectation
value in the nucleon sector, fig. 1(a), would change. A part of this
change is through the change of wave functions. Another important
effect comes from the change of three-body correlations.
In the framework of the lowest order Brueckner theory \cite{LOBT1, LOBT2},
they are represented through the modification of Pauli effects and
starting energy dependence. Since the change of nucleon Pauli effects
is absent for the addition of hyperons, a relevant correlation is
the potential insertion for hole states, fig. 1(b). A requirement of
self-consistency \cite{BG} for the hole energies means the inclusion of
some second order hole-line insertions, figs. (c) and (d).
\begin{figure}[t]
\epsfxsize=10cm
\epsfbox{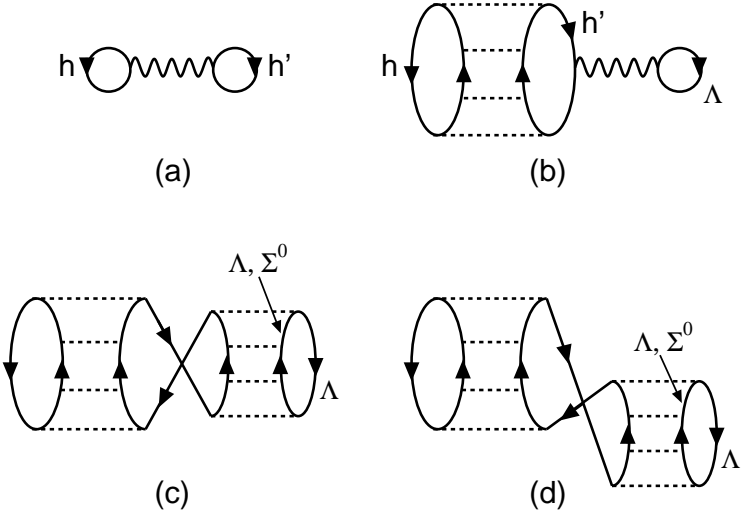}
\caption{Diagrams in the hole-line expansion in the Brueckner theory.}
\end{figure}

In the lowest order Brueckner theory, a ground state potential energy of
\he, starting from the realistic $NN$ interaction $v_{NN}$, is given by
introducing $G$-matrix elements:
\begin{eqnarray}
 PE_{\alpha}(4)&=&\frac{1}{2}\sum_{hh'}
 \langle hh' |G_{NN}(4,\omega =e_h(4) +e_{h'}(4))|hh'\rangle_{as}, \\
  e_h(4) &=& \langle h|t_N |h\rangle  +\sum_{h'}
   \langle hh' |G_{NN}(4,\omega =e_h(4) +e_{h'}(4))|hh'\rangle_{as}, \\
 G_{NN}(4,\omega) &=& v_{NN} +v_{NN}\frac{Q}{\omega - H_0}
G_{NN}(4,\omega ),
\end{eqnarray}
where $h$ and $h'$ correspond to a sole occupied $0s$ state besides
implicit spin and isospin summations.
The standard choice of the intermediate spectra $H_0$ is a $QTQ$
prescription. The self-consistency of $e_h$ and $G$ means that the hole
line potential insertion, a part of three-body correlations, is taken
into account. In finite nuclei Hartree-Fock condition is also required.
However, in the \he nucleus a single harmonic oscillator function
provides a very good approximation. It is also known that the \he core
in the \hl differs little from the \he. In the following discussion we
assume that the single-particle wave function is given by the same
harmonic oscillator (h.o.) (0s) function both
for \he and \hl. Then, the potential energy of the \he core in \hl
is also given by an expression similar to eq. (1):

\begin{eqnarray}
 PE_{\alpha}(5)&=&\frac{1}{2}\sum_{hh'}
 \langle hh' |G_{NN}(5,\omega =e_h(5) +e_{h'}(5))|hh'\rangle_{as}, \\
 e_h(5) &=& \langle h|t_N |h\rangle +\sum_{h'}
 \langle hh' |G_{NN}(5,\omega =e_h(5) +e_{h'}(5))|hh'\rangle_{as} \nonumber \\
 & & + \langle h\Lambda |G_{N\Lambda}(5,\omega =e_h(5)
  +e_\Lambda (5))|h\Lambda \rangle , \\
 G_{NN}(5,\omega) &=& v_{NN} +v_{NN}\frac{Q}{\omega - H_0}
 G_{NN}(5,\omega ).
\end{eqnarray}
In this case the single-particle potential includes a contribution of the
$\Lambda$ particle. The added lambda particle does not modify Pauli exclusion
operators for nucleons in eqs. (3) and (6). In the above framework
the difference between $PE_\alpha (4)$ and $PE_\alpha (5)$ comes from the
difference of $e_h$.

To estimate the difference between $PE_\alpha (5)$ and $PE_\alpha (4)$,
we rewrite the expression as follows:
\begin{equation}
 \Delta PE \equiv PE_\alpha (5)-PE_\alpha (4)
 = \frac{1}{2}\sum_{hh'} \langle hh' |G_{NN}(5)-G_{NN}(4)|hh'\rangle_{as}.
\end{equation}
Using the well-known relation
\begin{eqnarray}
  & & \langle hh'| G_{NN}(5)-G_{NN}(4)|hh'\rangle_{as} \nonumber \\
 &=&\langle hh'| G_{NN}(4)\left\{ \frac{Q}{e_h(5)+e_{h'}(5) -QTQ}
  - \frac{Q}{e_h(4)+e_{h'}(4) -QTQ}
  \right\} G_{NN}(5) |hh'\rangle_{as} \nonumber \\
 &=& \langle hh'| G_{NN}(4) \frac{Q}{e_h(4)+e_{h'}(4) -QTQ}
 (e_h(4)+e_{h'}(4) -e_h(5)-e_{h'}(5)) \nonumber \\
 & & \hspace*{5cm} \times \frac{Q}{e_h(5)+e_{h'}(5)-QTQ} G_{NN}(5)|hh'\rangle_{as},
\end{eqnarray}
$\Delta PE$ is expressed as
\begin{equation}
 \Delta PE = - \frac{1}{2} \sum_{hh'} \Delta e_{h} \langle hh'|
 G_{NN}(4)  \frac{Q}{e_h(4)+e_{h'}(4) -QTQ}
  \frac{Q}{e_h(5)+e_{h'}(5) -QTQ} G_{NN}(5) |hh'\rangle_{as} ,
\end{equation}
where $\Delta e_h$ is
\begin{equation}
 \Delta e_h \equiv  e_h(5) +e_{h'}(5) -e_h(4)-e_{h'}(4) =2( e_h(5) -e_h(4) ).
\end{equation}
At this stage we may introduce an approximation of
$G_{NN}(5)\sim G_{NN}(4)$ and $e_h(5)\sim e_h(4)$ in the denominator.
Using the relation
\begin{equation}
 \frac{\partial G_{NN}}{\partial \omega} = -G_{NN}\frac{Q}{\omega -QTQ}
 \frac{Q}{\omega -QTQ}G_{NN},
\end{equation}
the difference of the potential energy expectation value becomes
\begin{equation}
 \Delta PE \simeq \frac{1}{2} \sum_{hh'} \Delta e_h \langle hh'|
 \frac{\partial G_{NN}}{\partial \omega} |hh'\rangle_{as} = -\frac{1}{2}
  \sum_{h} \Delta e_h \kappa_N,
\end{equation}
where we introduced $\kappa_N \equiv - \sum_{h'}
 \langle hh'|\frac{\partial G_{NN}}{\partial\omega}|hh'\rangle_{as}$ which has been
known as a wound integral.

There being only one single-particle state,
$\Delta e_h$ is also expressed in terms of
 $\frac{\partial G_{NN}}{\partial \omega}$:
\begin{eqnarray}
 \Delta e_h &=&2(e_h (5)- e_h (4)) \nonumber \\
 &=&  2 \langle h\Lambda |G_{N\Lambda}(5)|h\Lambda \rangle
 + 2 \sum_{h'} \langle hh'|G_{NN}(5)-G_{NN}(4)|hh'\rangle_{as} \nonumber \\
 &\simeq& 2 \langle h\Lambda |G_{N\Lambda}(5)|h\Lambda \rangle - 2 \Delta e_h
  \kappa_N .
\end{eqnarray}
Thus we find
\begin{equation}
 \Delta e_h \simeq \frac{2}{1+2\kappa_N}
 \langle h\Lambda |G_{N\Lambda}(5)|h\Lambda \rangle.
\end{equation}
Substituting this result for $\Delta e_h$ in eq. (12),
we finally obtain the following expression.
\begin{equation}
 \Delta PE \simeq -\sum_h
  \frac{\kappa_N}{1+2\kappa_N}
 \langle h\Lambda |G_{N\Lambda}(5)|h\Lambda \rangle.
\end{equation}

Estimation of the matrix element
$\langle h\Lambda |G_{N\Lambda}|h\Lambda \rangle$
requires a knowledge of the $\Lambda$ and nucleon wave functions.
However, this matrix element can be related to the $\Lambda$ separation
energy $\epsilon_\Lambda$ which is known experimentally to be $-3.12$ MeV;
\begin{equation}
 \epsilon_\Lambda = E(\mbox{\hl})-E(\mbox{\he})
 = \langle \Lambda |t_\Lambda |\Lambda \rangle
  + \sum_h \langle h\Lambda |G_{N\Lambda}(5) |h\Lambda \rangle
  + \Delta PE +\Delta T_{cm},
\end{equation}
where we write the difference of the center of mass kinetic energy
as $\Delta T_{cm}$.
In order to simplify expressions we utilize the fact that there is
only one nucleon single-particle state. Inserting $\Delta PE$ of eq. (15)
into the right hand side, we obtain
\begin{equation}
 \epsilon_\Lambda (5)-\langle \Lambda |t_\Lambda |\Lambda \rangle
  -\Delta T_{cm} \simeq \sum_h \frac{1+\kappa_N}{1+2\kappa_N}
 \langle h\Lambda |G_{\Lambda N}(5)|h\Lambda \rangle.
\end{equation}
Hence, eliminating the $\Lambda N$ matrix element in eq. (14), we end up
with the following estimation:
\begin{equation}
 \Delta PE \simeq -\frac{\kappa_N}{1+\kappa_N}
 (\epsilon_\Lambda (5)-\langle \Lambda |t|\Lambda\rangle -\Delta T_{cm}).
\end{equation}
It is worthwhile to note that this difference of the potential
energy contributions is due to the effect through the
starting energy dependence of $NN$ $G$-matrices. The addition of
the $\Lambda$ to \he makes the single-particle energy of the nucleon
deeper, which induces less attractive $NN$ reaction matrices.
Thus, the expression of eq. (18) does not explicitly include the
quantity such as $\Lambda N$ correlations.

Supposing a single harmonic oscillator wave function with the
oscillator constant $\nu_\Lambda$, the kinetic energy expectation
value $\langle \Lambda\! |t_\Lambda |\Lambda\! \rangle$ is expressed as
$\frac{3}{4}\frac{\hbar^2}{m_\Lambda}\nu_\Lambda$, while $\Delta T_{cm}$
is given by $\frac{3}{4}\frac{\hbar^2}{m} (\frac{m_\Lambda}{4m+m_\Lambda}
\nu -\frac{m}{4m+m_\Lambda} \nu_\Lambda )$ with $m$ and $m_\Lambda$
being the nucleon and lambda masses, respectively.
The wound integral $\kappa_N$ is estimated by employing nuclear matter
$G$-matrix calculations in \he in the following scheme. The matrix
element $\langle 00(\frac{\nu}{2})| V_{NN}|00(\frac{\nu}{2})\rangle$
between $0s$ nucleon states with
the oscillator constant $\nu$ is evaluated by
\begin{equation}
  \langle 00(\frac{\nu}{2})| V_{NN}|00(\frac{\nu}{2})\rangle
   =\left( \frac{8\pi}{\nu}\right)^{\frac{3}{2}}
  \frac{1}{(2\pi^2)^2} \int_0^\infty dq' \int_0^\infty dq q'^2 q^2
  \mbox{e}^{-\frac{1}{\nu}(q^2 +q'^2)} \langle
  q'|G_{NN}^{\ell =0}(k_F)|q\rangle ,
\end{equation}
where $\langle q'|G_{NN}^{\ell =0}(k_F)|q\rangle$ is obtained by
the equation
\begin{equation}
 \langle q'|G_{NN}(k_F)|q\rangle =\langle q'|v_{NN}|q\rangle
  +\langle q'|v_{NN}\frac{Q(k_F)}{\omega -QTQ}G_{NN}(k_F)|q\rangle .
\end{equation}
In numerical calculations, a standard oscillator constant
of $\nu=0.56$ fm$^{-2}$ was taken for \he, and the Fermi momentum $k_F$
was set to be 1.2 fm$^{-1}$ since the average density $\bar{\rho}
\equiv \int \{\rho (r)\}^2 r^2 dr/\int \rho(r) r^2 dr$ is 0.106 fm$^{-3}$
which corresponds to $k_F \sim 1.2$  fm$^{-1}$.
The energy dependence of $\langle q'|G^{\ell =0}(k_F)|q\rangle$
tells us that $\kappa_N$ is about $0.2$, which is reasonable.
Expecting $\nu_\Lambda$ to be $0.4 \sim 0.5$ fm$^{-2}$,
we obtain $\Delta PE$ as $2.5 \sim 2.9$ MeV. Although further
contributions from orbital rearrangement and other higher order
correlations are expected, it is important to settle the order of
magnitude of rearrangement effects in \hl by simple and transparent
arguments.

\section{Rearrangement effects in \dll of \hll}

In this section, we consider the energy
$\Delta B_{\Lambda\Lambda} \equiv 2E(\mbox{\hl})
-E(\mbox{\hll})-E(\mbox{\he})$. It is straightforward
to decompose it to each matrix element:
\begin{eqnarray}
 \Delta B_{\Lambda\Lambda} &=& \frac{1}{2} \sum_{hh'}
 \langle hh'|2G_{NN}(5)-G_{NN}(4)-G_{NN}(6) |hh'\rangle_{as} \nonumber \\
 &+& 2\sum_h \langle h\Lambda | G_{N\Lambda}(5) - G_{N\Lambda}(6)
 |h\Lambda \rangle - \langle \Lambda\Lambda | G_{\Lambda\Lambda}(6)
 |\Lambda\Lambda \rangle_{as} +\Delta T_{\Lambda\Lambda} ,
\end{eqnarray}
where the single-particle wave functions are assumed to be common
in \he, \hl and \hll, and $\Delta T_{\Lambda\Lambda}$ is the contribution
of kinetic energy terms which is discussed later.
In order to obtain an information for the strength of the $\Lambda\Lambda$
interaction $\langle \Lambda\Lambda | G_{\Lambda\Lambda}(6) |
 \Lambda\Lambda \rangle_{as}$
from experimental data of $E(\mbox{\hl})$, $E(\mbox{\hll})$ and
$E(\mbox{\he})$, it is necessary to estimate rearrangement contributions
which correspond to the first and the second terms of the above expression.
Since the changes from $G_{NN}(4)$ to $G_{NN}(5)$ and from
$G_{NN}(5)$ to $G_{NN}(6)$ are the same in the leading order, the first
term of eq. (21) should be small because of the cancellation
$(G_{NN}(5)-G_{NN}(4))-(G_{NN}(6)-G_{NN}(5)) \sim 0$.
The second term represents the main source of the rearrangement effect.
As in eq. (8), it is straightforward to obtain
\begin{eqnarray}
 \langle h\Lambda| G_{N\Lambda}(6)&-&G_{N\Lambda}(5)|h\Lambda \rangle =
 \langle h\Lambda|G_{N\Lambda}(5) \left\{
  \frac{Q(6)}{e_h(6)+e_{\Lambda}(6)-Q(6)TQ(6)} \right. \nonumber \\
 & &\hspace*{22mm} -\left. \frac{Q(5)}{e_h(5)+e_{\Lambda}(5) -Q(5)TQ(5)}
  \right\} G_{N\Lambda}(6) |h\Lambda\rangle \nonumber \\
 &=& \langle h\Lambda|G_{N\Lambda}(5)
 \frac{Q(6)-Q(5)}{e_h(5)+e_{\Lambda}(5)-Q(5)TQ(5)}
  G_{N\Lambda}(6)|h\Lambda\rangle \nonumber \\
 &-& \langle h\Lambda|G_{N\Lambda}(5)
 \frac{\Delta e_{h\Lambda}Q(6)-(Q(6)-Q(5))TQ(6)}
 {(e_h(5)+e_{\Lambda}(5) -Q(5)TQ(5))(e_h(6)+e_{\Lambda}(6) -Q(6)TQ(6))}
   G_{N\Lambda}(6) |h\Lambda\rangle ,
\end{eqnarray}
where $\Delta e_{h\Lambda}\equiv e_h(6)+e_\Lambda(6)-e_h(5)-e_{\Lambda}(5)$
is a single-particle energy difference. In this case the change of
the Pauli blocking for $\Lambda$ also contributes. In the following
presentation, we neglect the contribution from the term including
$(Q(6)-Q(5))TQ(6)$ because of its restricted summation compared with
$\Delta e_{h\Lambda}Q(6)$.

Writing $Q(5)-Q(6)$ as $\sum_p |p\Lambda_0 \rangle\langle p\Lambda_0|$
where $\Lambda_0$ stands for a $\Lambda$ $0s$ state and $|p\rangle$
for a nucleon unoccupied state,
\begin{eqnarray}
 \langle h\Lambda | G_{N\Lambda}(6)-G_{N\Lambda}(5)|h\Lambda \rangle &=&
 -\sum_p \frac{\langle h\Lambda|G_{N\Lambda}(5)|p\Lambda_0 \rangle
 \langle p\Lambda_0 | G_{N\Lambda}(6)|h\Lambda \rangle}
 {e_h(5)+e_\Lambda (5)-Q(5)TQ(5)}
  \nonumber \\
   & &+ \Delta e_{h\Lambda}\langle h\Lambda |
   \frac{\partial G_{N\Lambda}(5)}{\partial \omega} |h\Lambda \rangle .
\end{eqnarray}
The first term is positive because the numerator is positive and the
denominator is negative. The second term is also positive
if $\Delta e_{h\Lambda} < 0$, as the $\Lambda N$ interaction
$\langle h\Lambda | G_{N\Lambda} (5)|h\Lambda \rangle$ becomes more
attractive when the starting energy $\omega$ becomes shallower and
thus the derivative of the matrix element with respect to $\omega$
is negative. To estimate $\Delta e_{h\Lambda}$, we calculate the
energy differences $e_h (6)- e_h (5)$ and $e_\Lambda (6)-e_\Lambda (5)$
as follows:
\begin{eqnarray}
 e_h(6)-e_h(5)&=& \sum_{h'} \langle  hh'|
 \frac{\partial G_{NN}}{\partial \omega} |hh'\rangle
 \Delta e_h+\langle h\Lambda |
 \frac{\partial G_{N\Lambda}}{\partial\omega}|h\Lambda \rangle
 \Delta e_{h\Lambda}
 \nonumber \\
 &-& \sum_p \frac{\langle h\Lambda |G_{N\Lambda}(5)|p\Lambda_0 \rangle
 \langle p\Lambda_0 |G_{N\Lambda}(6)|h\Lambda \rangle}
 {e_h(5)+e_\Lambda (5)-Q(5)TQ(5)}
 +\langle h\Lambda | G_{N\Lambda}(6)|h\Lambda \rangle , \\
e_\Lambda (6)-e_\Lambda (5)&=& \sum_{h'}
 \langle \Lambda h'|\frac{\partial G_{\Lambda N}}{\partial \omega}
  |\Lambda h'\rangle
 \Delta e_{h' \Lambda} \nonumber \\
 &-& \sum_{h'p} \frac{\langle \Lambda h'|G_{\Lambda N}(5)|\Lambda_0 p
 \rangle \langle \Lambda_0 p|G_{\Lambda N}(6)|\Lambda h'\rangle}
 {e_{h'}(5) +e_\Lambda (5)-Q(5)TQ(5)}
 +\langle \Lambda\Lambda | G_{\Lambda\Lambda}(6)|\Lambda\Lambda \rangle .
\end{eqnarray}
In order to simplify these expressions, we introduce following notations:
\begin{eqnarray}
 W&=& \sum_{h'}
 \langle \Lambda h'|\frac{\partial G_{\Lambda N}}{\partial \omega} |
 \Lambda h'\rangle \Delta e_{h' \Lambda}
 = -\Delta e_{h\Lambda} \kappa_\Lambda , \\
 P&=& - \sum_{h'p} \frac{\langle \Lambda h'|G_{\Lambda N}(5)|
 \Lambda_0 p\rangle \langle \Lambda_0 p|G_{\Lambda N}(6)|
 \Lambda h'\rangle}{e_{h'}(5)+e_\Lambda (5)-Q(5)TQ(5)},
 \\
 D&=&\sum_h \langle h\Lambda | G_{N\Lambda}(6)-G_{N\Lambda}(5)|
 h\Lambda \rangle = W+P.
\end{eqnarray}
Using the wound integral $\kappa_\Lambda = -\sum_{h'}
 \langle \Lambda h'|\frac{\partial G_{\Lambda N}}{\partial \omega} |
 \Lambda h'\rangle$
for the $\Lambda N$ pair, the energy differences are written as
\begin{eqnarray}
 e_h(6)-e_h(5)&=& -\Delta e_h \kappa_N +\frac{1}{4} D
  +\langle h\Lambda | G_{N\Lambda}(6) | h\Lambda \rangle
  = -\Delta e_h \kappa_N +\frac{1}{2} D+\langle h\Lambda |
  G_{N\Lambda}(5) | h\Lambda \rangle , \\
 e_\Lambda (6)-e_\Lambda (5)&=& D
 +\langle \Lambda\Lambda | G_{\Lambda\Lambda}(6)|\Lambda\Lambda \rangle .
\end{eqnarray}
Then, noticing the relation $e_h(6)-e_h(5)=\frac{1}{2}\Delta e_h$,
$\Delta e_{h\Lambda}= e_h(6) -e_h(5)+e_\Lambda (6)-e_\Lambda (5)$ becomes
\begin{equation}
 \Delta e_{h\Lambda}=\frac{3+4\kappa_N}{2(1+2\kappa_N )}D+\frac{1}{1+2\kappa_N}
 \langle h\Lambda | G_{N\Lambda}(5) | h\Lambda \rangle
  +\langle \Lambda\Lambda | G_{\Lambda\Lambda}(6) |\Lambda\Lambda \rangle .
\end{equation}
Inserting this into $D=W+P=-\kappa_\Lambda \Delta e_{h\Lambda} +P$, $D$ can be
expressed as
\begin{equation}
 D= \frac{1}{1+\kappa_\Lambda \frac{3+4\kappa_N}{2(1+2\kappa_N )}} \left(
  \frac{-\kappa_\Lambda}{1+2\kappa_N} \langle h\Lambda |
  G_{N\Lambda}(5)|h\Lambda \rangle -\kappa_{\Lambda}\langle \Lambda\Lambda |
  G_{\Lambda\Lambda}(6)|\Lambda\Lambda \rangle +P\right).
\end{equation}
The estimation of $\Delta B_{\Lambda\Lambda}$ finally reads
\begin{equation}
\Delta B_{\Lambda\Lambda} \sim -\langle \Lambda\Lambda | G_{\Lambda\Lambda}(6)
 |\Lambda\Lambda \rangle  -2D +\Delta T_{\Lambda\Lambda}.
\end{equation}
Since $D$ is positive as remarked above, the $\Delta B_{\Lambda\Lambda}$
becomes smaller than $-\langle \Lambda\Lambda | G_{\Lambda\Lambda}(6)|
\Lambda\Lambda \rangle$. Some comments are necessary for the contribution
of $\Delta T_{\Lambda\Lambda}$. The assumption of the same single-particle
wave functions implies
\begin{equation}
\Delta T_{\Lambda\Lambda}=\frac{3\hbar^2}{4m}\nu \frac{2m_\Lambda^2}
{(4m+m_\Lambda )(4m+2m_\Lambda )}\left( 1-\frac{m}{m_\Lambda}
\frac{\nu_\Lambda}{\nu}\right) ,
\end{equation}
which amounts to 0.37 MeV with $\nu=0.56$
fm$^{-2}$ and $\nu_\Lambda = 0.5$ fm$^{-2}$, and thus partly cancels
the negative contribution of $-2D$. However, $\Delta T_{\Lambda\Lambda}$ is
sensitive to the change of single-particle wave functions. It has been known
that the addition of the $\Lambda$ to \hl makes the $\Lambda$ single-particle
wave function compact \cite{TXB}. In that case, $\Delta T_{\Lambda\Lambda}$
becomes even negative. In the estimation below we leave out this contribution.

The matrix element  $\langle h\Lambda | G_{N\Lambda}(5)| h\Lambda \rangle$
in eq. (32) is estimated by eq. (17). New ingredients here are the wound
integral $\kappa_\Lambda$ and the Pauli blocking effect $P$. The energy
dependence of the $\Lambda N$ reaction matrix element is much weaker than
that of the $NN$ interaction, since the $\Lambda N$ tensor force, which
is an important source of the $\omega$-dependence, is weak because of
the absence of the lowest order pion exchange.
We find $\kappa_\Lambda \sim 0.05$ from the $\omega$-dependence of
the calculated $\Lambda N$ reaction matrices. The Pauli
blocking effect is estimated by evaluating the matrix element
$\langle 00(\nu )|V_{\Lambda N}|00(\nu )\rangle$ from the hypernuclear
matter $G$ matrix $\langle q'|G_{\Lambda N}(k_F^N ,k_F^\Lambda )|q\rangle$
by solving the equation similar to eq. (20). Half of the difference between
matrix elements with
$(k_F^N ,k_F^\Lambda )= (1.2\mbox{ fm}^{-1}, 0 \mbox{ fm}^{-1})$ and
$(k_F^N ,k_F^\Lambda )= (1.2\mbox{ fm}^{-1}, 1.2\mbox{ fm}^{-1})$
gives $P \sim 0.23$ MeV.
The calculation of $\langle \Lambda\Lambda | G_{\Lambda\Lambda} |
\Lambda\Lambda \rangle$ is carried out, as in eq. (19), by the momentum
space folding of nuclear matter $\Lambda\Lambda$ $G$-matrix, in which
the coupling with the $\Sigma\Sigma$ and $\Xi N$ channels is included.
The quark model potential fss2 \cite{FU01} indicates that
$\langle \Lambda\Lambda | G_{\Lambda\Lambda} |\Lambda\Lambda \rangle$
is $-2.17$ MeV for $\nu = 0.40$ fm$^{-2}$,
and $-2.65$ MeV for $\nu = 0.50$ fm$^{-2}$. In these
calculations, the starting energy $\omega=2e_\Lambda (6)$ in the energy
denominator of the $G$-matrix equation is set to be $-12$ MeV.
The $\kappa_\Lambda$ is 0.05 for $\nu = 0.40$ fm$^{-2}$,
and 0.06 MeV for $\nu = 0.50$ fm$^{-2}$. Then we obtain $2D=0.93$ MeV,
namely \dll = 1.24 MeV, for $\nu = 0.40$ fm$^{-2}$ and $2D=1.12$ MeV,
namely \dll = 1.53 MeV, for $\nu = 0.50$ fm$^{-2}$.
Since we should expect other effects not considered here such as the
change of wave functions and the contribution of the first term
of eq. (21), more quantitative evaluation of the actual rearrangement
contribution would be desirable.

It is instructive to present the matrix element
$\langle \Lambda\Lambda | G_{\Lambda\Lambda} |\Lambda\Lambda \rangle$
calculated with other available $\Lambda\Lambda$ interactions.
Lanskoy and Yamamoto \cite{LY} parameterized $k_F^N$-dependent
$\Lambda\Lambda$ $G$-matrices \cite{YMHIN} in a three-range Gaussian
form obtained from the $\Lambda\Lambda$-$\Xi N$ sectors of the
Nijmegen hard core model-D (ND) \cite{NIJDF} and the Nijmegen soft core
model (NS) \cite{NIJNS}. Nishizaki, Yamamoto and Takatsuka \cite{NYT}
also gave parameters of the $k_F^N$-dependent effective $\Lambda\Lambda$
interactions in a four-range Gaussian form, based on $G$-matrices
of the Nijmegen model-D and model-F (NF) potentials \cite{NIJDF}. The
obtained matrix elements with these effective forces at $k_F^N=1.2$ fm$^{-1}$
are shown and compared with our results of the quark model potential
fss2 \cite{FU01} in Tables I. Referring to the newly determined value
of \dll (\hll ) $\sim 1$ MeV \cite{TAKA}, the Nijmegen model $D$ gives
stronger attraction, though there is ambiguity in the choice of the hard
core radius. On the other hand, the Nijmegen soft core model has an
insufficient $\Lambda\Lambda$ attraction,
when we take into account the Brueckner rearrangement effects.
Vida\~{n}a {\it et al.} \cite{VPRS} calculated bond energies \dll
for heavier double-lambda hypernuclei by a new set of the Nijmegen
soft core potentials \cite{RSY}. Their results suggested that the
new soft core potentials have a weaker $\Lambda\Lambda$ attraction.

\begin{table}
\setlength{\tabcolsep}{0.2in}
\parbox{0.9\textwidth}{\caption{Matrix elements
$\langle \Lambda\Lambda | G_{\Lambda\Lambda} |\Lambda\Lambda \rangle$
calculated by $k_F^N$-dependent effective $\Lambda\Lambda$ interactions
parameterized by Lanskoy and Yamamoto (LY) \cite{LY},
and Nishizaki, Yamamoto and Takatsuka (NYT) \cite{NYT},
which are based on $G$-matrices obtained from three models
by the Nijmegen group \cite{NIJDF,NIJNS}: model-D (ND), model-F (NF) and
the soft core model (NS). Our results of the quark model potential fss2
\cite{FU01} are also included. Calculations are done for two
harmonic oscillator constants of the $\Lambda$ single-particle wave function.
$k_F^N$ is set to be 1.2 fm$^{-1}$. Entries are in MeV.\\}}

\begin{tabular}{rccccc} \hline\hline
                     & LY-ND & LY-NS & NYT-ND & NYT-NF & fss2 \\ \hline
 $\nu=0.4$ fm$^{-2}$ & -4.03 & -1.27 & -3.68 & +2.68 & -2.17 \\
 $\nu=0.5$ fm$^{-2}$ & -4.90 & -1.57 & -4.29 & +3.90 & -2.65 \\ \hline \hline
\end{tabular}
\end{table}

\section{Conclusion}
We have estimated rearrangement contributions in the energy expectation values
of the \hl and \hll systems. The knowledge of
this quantity is important to deduce the information of strengths of
hyperon-nucleon and hyperon-hyperon interactions from experimental
binding energies. In the standard framework of the lowest order Brueckner
theory \cite{LOBT1,LOBT2}, from which invaluable understanding of nuclear
systems has been accumulated for more than 40 years, the principal
contribution to rearrangement energies is due to the starting energy
dependence of the effective interaction as well as the change of
Pauli blocking effect. Since we have avoided specific numerical
calculations as much as possible, our results may not be very quantitative.
On the other hand, the magnitude of the estimated values is rather
robust. The energy of the $\alpha$ core part is reduced by about 2 MeV
when the single-particle energy becomes deeper by the addition of $\Lambda$ to
\he . This effect is connected with the overbinding problem of the $s$-shell
hypernulcei, as Bando and Shimodaya \cite{BS80} discussed in 1980.
The new aspect of our presentation is the treatment of the estimation
of the matrix element $\langle \Lambda h |G_{\Lambda N}|\Lambda h\rangle$.
Instead of calculating it directly, we relate it to the $\Lambda$ separation
energy, which is by itself influenced by the rearrangement energy.
As it should be, our result confirmed their estimation.

The rearrangement energy discussed in this paper is certainly not a sole
answer for the overbinding problem. Core polarization effects should be
considered, though the \he core is rather rigid. Several estimations
\cite{BU,BMU,LAN} suggested that the energy change due to the core
polarization is about 0.5 MeV or less. Three-body correlations and
$\Lambda NN$ three-body forces should also be taken into account.
In the former case the coupling to the $\Sigma$
intermediate state plays a characteristic role. It is important
to treat three-body correlation effects together without three-body
forces, which is an interesting subject for future investigation.

We next considered rearrangement effects to obtain the matrix element
of the $\Lambda\Lambda$ interaction from the observed masses of
double lambda hypernuclei. The experimental determination of the
\dll which is defined as $2E(\mbox{\hl})-M(\mbox{\he})
-M(\mbox{\hll})$ does not directly tell the strength of
the $\Lambda\Lambda$ interaction, since it contains the
rearrangement energies. This investigation is important in view of the
recent experimental finding, which would renew previous understanding that
the $\Lambda \Lambda$ interaction is rather strong. We derived a compact
expression for the \dll, assuming common wave functions for \he, \hl and \hll.
The present estimation of the rearrangement contribution is about 1 MeV.
Thus the $s$-wave matrix element of the $\Lambda\Lambda$ interaction
in \hll should be around $-2$ MeV.

Our calculation is based on the argument of the many body theory in the
model space. The extension of few body type calculations using bare
interactions for hypernuclei such as \hl and \hll would clarify the same
subject directly from the basic microscopic viewpoint. As for the \hl,
Nemura, Akaishi and Suzuki \cite{NAS} recently showed by variational
calculations that \he core energy in \hl
decreases by 4.7 MeV. The difference of about 2 MeV may come from actual
change of wave functions and higher order effects. It is interesting and
gratifying to observe the correspondence between these studies and our
treatments in the standard many-body theory.

Finally we comment that similar considerations should be applied to
$^9_\Lambda$Be and $^{\hspace*{0.2em}10}_{\Lambda\Lambda}$Be, where
ad hoc effective interactions tend to have been employed.

\end{document}